# Selection of Arginine-Rich Anti-Gold Antibodies Engineered for Plasmonic Colloid Self-Assembly

*Purvi Jain[1], Anandakumar Soshee[1], S Shankara Narayanan[3], Jadab Sharma[3], Christian Girard[3], Erik Dujardin[3*], Clément Nizak[1,2*]*

[1] Laboratory of Interdisciplinary Physics, UMR5588 Grenoble Université 1/CNRS, Grenoble, France.

[2] Laboratory of Biochemistry, UMR 8231 ESPCI ParisTech/CNRS, Paris, France.

[3] CEMES CNRS UPR 8011, 29 rue J. Marvig, 31055 Toulouse Cedex 4, France.

**KEYWORDS.** Antibody, phage display, arginine, gold-affinity interactions, plasmonic nanoparticles, self-assembly, coupled local surface plasmons.

**ABSTRACT.** Antibodies are affinity proteins with a wide spectrum of applications in analytical and therapeutic biology. Proteins showing specific recognition for a chosen molecular target can be isolated and their encoding sequence identified *in vitro* from a large and diverse library by phage display selection. In this work, we show that this standard biochemical technique rapidly yields a collection of antibody protein binders for an inorganic target of major technological importance: crystalline metallic gold surfaces. 21 distinct anti-gold antibody proteins emerged from a large random library of antibodies and were sequenced. The systematic statistical analysis of all the protein sequences reveals a strong occurrence of arginine in anti-gold antibodies, which corroborates recent molecular dynamics predictions on the crucial role of arginine in protein/gold interactions. Once tethered to small gold nanoparticles using histidine tag chemistry, the selected antibodies could drive the self-assembly of the colloids onto the surface of single crystalline gold platelets as a first step towards programmable protein-driven construction of complex plasmonic architectures. Electrodynamic simulations based on the Green Dyadic Method suggest that the antibody-driven assembly demonstrated here could be exploited to significantly modify the plasmonic modal properties of the gold platelets. Our work shows that molecular biology tools can be used to design the interaction between fully folded proteins and inorganic surfaces with potential applications in the bottom-up construction of plasmonic hybrid nanomaterials.

Taking advantage of the interaction specificity and engineering versatility of biomolecules holds promises towards the design of hybrid self-assembled materials by combining bio-molecules and non-biological solid state nano-objects [1]. For instance DNA has been used to self-assemble nano-particles into three-dimensional (3D) crystals[2,3]. Proteins and peptides represent an alternative strategy offering a much wider chemical (20 amino-acids vs 4 nucleotides) and structural (viral capsids, microtubules, S-layers, …) repertoire. However, this strategy requires a deeper understanding of interactions between proteins and non-biological solid surfaces, which remains essentially elusive. Combinatorial chemistry has already demonstrated its power in revealing property-driven best-suited molecules [4], materials [5] or biomolecules [6] with no *a priori* knowledge of interaction mechanisms between the constituents and the environment in which the selection pressure is applied. Biomolecular display techniques are among the most advanced tools to date developed for such directed evolution approaches in biochemistry. Interestingly, peptides that bind tightly and specifically to solid surfaces such as metals, semi-conductors, magnetic metal oxides, conductive polymers and carbon nanotubes have been identified by *in vitro* phage-display screening of libraries containing typically $10^9$ distinct, linear 9-12 amino-acid peptides [7-9]. Thereby, a better description of interactions between these peptides and their inorganic target surfaces is emerging through combined AFM, NMR and peptide sequence analysis [10-13]. Remarkable advances in the affinity chemistry of biomolecule for solid interfaces have led to the programmed self-assembly of new materials based on phage bearing these peptides and their inorganic targets [14,15]. To overcome intrinsic limitations of peptides, in particular the lack of a well-defined 3D structure, antibodies have been proposed as stronger and more stable binders [16]. Therapeutic or cell-biological applications of antibodies rely on libraries of $10^9$ recombinant antibody clones, that mimic human antibody repertoires, and which are typically screened by phage display for isolating the few antibody interactants binding to the chosen bio-molecular targets [6]. Such libraries have also been screened against inorganic solid surfaces, such as metals (Au) [17] and semi-conductors (GaAs) [16]. These screens yield highly specific, versatile antibody binders of these inorganic crystalline solid surfaces, thus confirming the validity of the antibody-based combinatorial approach. Anti-Au antibodies do not bind other tested bare metal surfaces (Pt, Pd, Ag), and can be engineered to functionalize gold surfaces [17]. Anti-GaAs antibodies bind specifically to [111A] crystal facets and not to [100] facets [16]. While these seminal works highlight the potential of antibody-based nanomaterials engineering, they remain isolated cases, which makes it difficult to infer a more generic rationale.

In this work, a library of $3 \times 10^8$ distinct antibody clones was screened against bare gold surfaces of micron-sized particles and yielded a collection of more than 20 gold-binding antibodies. We shed a new light on the interactions between anti-Au antibodies and gold surfaces by comparing their sequences to those of antibodies isolated against a polymer target or to the sequence of antibodies randomly picked from the same initial library. Furthermore, we exploit the specific binding affinity of the selected anti-Au antibodies to decorate the bare gold surface of large crystalline gold nanoprisms with sub-5-nm Au nanoparticles tethered to the anti-Au antibodies. Atomic Force Microscope (AFM) and Transmission Electron Microscopy (TEM) imaging reveals the binding of the 5-nm Au nanoparticle across the entire prism edges and surface. We corroborate these findings by performing similar experiments with PVP-coated Au nanoprisms and a different anti-PVP antibody recently identified [18] that was conjugated to the small Au nanoparticles. The effective selective binding affinity of these antibodies even after grafting onto the nanoparticle surface is an important milestone towards protein-driven programmable self-assembly of functional nanomaterials such as plasmonic nanocrystals [19]. Green Dyadic simulations indeed show that the surface plasmon local density of states (SP-LDOS) of the Au nanoprism resonators is significantly modified by the presence of assembled nanoparticles on the nanoprism surface.

**Results and Discussion**

The Tomlinson I+J scFv antibody library contains $3 \times 10^8$ distinct antibody clones that share the same antibody framework and differ at 18 randomized amino-acid positions in the CDR2 and CDR3 regions of the heavy and light antibody Fv chains (respectively $V_H$ and $V_L$), responsible for specific interactions of antibodies with antigens (Figs. 1a,b). This widely distributed library has been extensively screened against bio-molecular targets for the past decade and yielded many antibodies. In two earlier studies, we have demonstrated that the interactions between antibodies and non-biological targets could also be analyzed and optimized by screening the Tomlinson I+J scFv library. Our general strategy has been applied to two distinct targets: (i) a single-stranded DNA [20,21] and (ii) poly-vinylpyrrolidone (PVP), a synthetic, neutral, water-soluble homopolymer of industrial interest [18]. In particular, a new amino-acid motif, that drives the specific recognition of PVP chains, has been isolated by antibody phage display selection followed by the statistical analysis of the amino-acid sequences of selected antibody clones and combined with biochemical characterization of antibody/target interactions. The PVP-binding motif comprises tyrosine, glycine, asparagine and aspartic acid residues and is located in the antigen-binding CDR3 region of the $V_H$ moiety [18].

Building up on our approach to analyze and exploit antibody/gold interactions, we have screened the Tomlinson I+J library against bare gold surfaces of micrometer-sized particles (Fig. 1c). The yield of phage recovery (output vs input phage particle numbers) is similar at the first two selection rounds ($10^{-7}$) but increases 500-fold at the third round ($5 \times 10^{-5}$). This suggests that three selection rounds result in the significant enrichment of the phage library with specific anti-Au antibodies. 94 antibody clones were randomly picked from the third selection round output population, arrayed in 96-well plates containing two negative control wells, and assayed by ELISA for binding to gold micron-sized particles. 45 antibody clones out of 94 tested (48%) did bind to gold particles. The antibody-Au binding specificity is confirmed by a series of negative control experiments. First, none of them bind to plain 96-well plate, therefore excluding plastic-bound phage selection (Fig. 2a). Secondly, selected anti-Au antibodies do not bind to bare metallic Ag, Cu or semiconducting ZnS surfaces, which rules out non-specific interactions of anti-Au antibodies with other inorganic solid surfaces (Fig. 2b). Thirdly, antibodies isolated from the same library, therefore sharing the same framework, but selected against PVP [18] do not bind to gold particles (Fig. 2c). This observation precludes non-specific interactions of the common antibody framework of the Tomlinson I+J library with gold surfaces. These experiments therefore demonstrate the specific binding of anti-Au antibodies to bare gold surfaces through their CDR regions. In contrast to previous reports of an anti-Au antibody that suffered from poor solubility [17], the anti-Au antibodies presented here are fully soluble and efficiently secreted by *E. coli* into the culture medium, with expression yields in the range from 1 mg to 10 mg per liter of culture.

In order to gain further insight on the molecular signature of Au binding affinity, 31 anti-Au antibody clones were sequenced leading to the identification of 21 distinct sequences, the most redundant clone being represented 8 times (See Supplementary Information). This large number of different antibodies showing gold affinity allows for a detailed statistical analysis of the protein sequences. To this end, the 21 anti-Au antibody sequences were aligned and a sequence logo analysis of the random CDR positions was performed as shown in Figure 3. The occurrence frequency of each of the 20 amino-acids is calculated for each random position, which yields a probability ranking represented by the symbol size in the sequence logo. For comparison, this analysis is applied to a set of 32 antibodies randomly picked from the initial library and to a set of 31 antibody selected from the same library against PVP [18]. The strong selection for arginine (Arg) residues in anti-Au antibodies is clearly visible at most of the random positions of CDR2 and CDR3 regions of $V_H$ and $V_L$. In contrast, Arg residues are either totally absent or, at most, present at up to one position in the initial population or in anti-PVP antibody clones. Noteworthy, amino-acids at random positions of the Tomlinson J library are NNK-encoded [22], which results in a bias for Arg residues (9%) with respect to most other amino-acids (3% for tyrosine, for example) [23]. This bias was, in part, compensated by using the I+J library that contains equal proportions of J library NNK-encoded clones and of I library DST-encoded clones, for which arginine residues are absent at random positions. Indeed, the maximum 1 Arg residue per antibody with 18 random positions found in the 32 random, non-selected clones from the initial I+J library corresponds to less than 6%. In contrast, in 21 anti-Au antibodies, 2 to 9 Arg residues are observed, which amounts to an average of $4.3 \pm 2$ Arg per antibody, that is $24 \pm 11\%$ of the random positions. This large occurrence of Arg residues reveals a statistically significant selection bias during the phage display rounds (Student's t test, $p=2.6 \times 10^{-6}$).

The involvement of Arginine in the Au surface binding is further confirmed by its spatial occurrence. Indeed, Arg is present in all CDR domains, which are the antigen-binding loops of antibodies. It is the 1$^{st}$ or 2$^{nd}$ top-ranked amino-acid at all 4 random positions of $V_H$-CDR3, which is considered as the most important CDR region for antigen binding. It holds the same predominance at 3 of the 7 random positions in the $V_H$-CDR2 domain, at 1 of the 2 random positions of $V_L$-CDR2 (3$^{rd}$ at the other position) and at 4 out of 5 random positions of $V_L$-CDR3. In this context, the only occurrence of Arg in the non-selected sequence logo is only ranked 4$^{th}$ at a single random position. Similarly, Arg is totally absent from all random positions of anti-PVP antibody clones. Besides the massively dominant role of Arg, Serine (Ser) and Glutamic acid (Glu) also emerge as statistically over-represented in anti-Au antibody sequences, where they are found in the $V_H$-CDR2 domain, albeit to a much lesser extent than Arg.

The observed predominance of Arg residues in our anti-Au antibodies is in agreement with recent molecular dynamics analysis on free amino-acids [24,25] or proteins containing β-sheet structures [26] interacting with [111] facets of crystalline gold surfaces. In these simulations [24,25], arginine is ranked respectively 1$^{st}$ and 6$^{th}$ out of the 20 free amino-acids, with an interaction energy as high as that of the sulfur-containing amino-acids, methionine and cysteine. Importantly, such thiolated residues are eliminated by our method due to the oxidizing conditions during antibody secretion by bacteria (see Experimental Section). In the case of β-sheet structures, time-lapsed simulations show that arginine residues play a crucial role in the initial contact with gold surfaces through interactions via their side chain [26].

Experimentally, arginine has been identified as a recurrent residue in the sequences of peptides that show strong affinity for gold surfaces. For example, arginine is the 3$^{rd}$ most represented amino-acid in gold-binding cyclic peptides isolated by bacterial cell-display. [11] Arginine also contributes to the decapeptide secreted by the bacterium *D. acetovorans* to precipitate Au nanoparticles as part of a protection mechanism against gold poisoning [27]. Interestingly, arginine does not appear in unstructured linear peptides isolated by phage-display against bare gold surfaces [28,29]. It is therefore possible that the interaction of arginine with gold surface may critically depend on the chemical or conformational context of a well-defined folded protein or constrained oligopeptides, while the ubiquitous cysteine and methionine affinity for gold rely primarily on the sulfur-gold iono-covalent bond formation.

Antibody-mediated interaction can be exploited to design complex self-assembly schemes in which surface affinity drives the construction of functional nanomaterials. To illustrate this bio-inspired self-assembly strategy, Au nanoparticles are derivatized with anti-Au antibodies, the gold affinity of which is subsequently used to drive the nanoparticle assembly predominantly onto the basal [111] facets of crystalline Au prisms (Fig. 4). First, the anti-Au antibodies that comprise a hexahistidine tag are coupled to 5-nm Ni-NTA-NanoGold at an equimolar ratio, using a standard protocol (See Experimental Section). PVP-coated Au nanoprisms synthesized as previously reported [30] are scattered on a clean glass substrate. Repeated oxygen plasma cycles are performed in order to remove the PVP capping layer while preserving the crystalline facets of the prisms (Figs. 4a-b). The cleaned prisms are then immediately incubated with a solution of antibody-conjugated Au nanoparticles followed by extensive rinsing with deionized water and blow dried (Fig. 4d). Atomic Force Microscopy (AFM) images of the resulting NanoGold-decorated Au nanoprisms show that the initially smooth prism surface, displayed in Figure 4b, is constellated with numerous protrusions of uniform height comprised between 3 and 5 nm (Fig. 4e and inset). Notably, very few nanoparticles are found on the silica substrate near the nanoprism, thus indicating that the antibody-functionalized NanoGold has a specific affinity for gold surfaces that results in an interaction strong enough to withstand the thorough rinsing step. Moreover, since the native surface of the prisms is coated with PVP, we could compare the affinity of anti-PVP and anti-Au antibodies for as-synthesized PVP-coated and plasma-cleaned Au nanoprisms. Hence, our protocol is directly applied to Au nanoprisms deposited onto a glass substrate but without performing the $O_2$ plasma step. The prisms are exposed to NanoGold functionalized with an anti-PVP antibody through a histidine / Ni-NTA coupling (Fig. 4d). Subsequent AFM images also demonstrate that nanoparticles of typically 3-5 nm height are attached specifically onto the PVP/Au surface as shown in Figure 4f. In the case of pristine prisms, we could also perform the self-assembly step in aqueous suspension and then examine the Au nanoprisms by Transmission Electron Microscopy, which confirms the attachment of anti-PVP NanoGold nanoparticle onto the surface of the PVP-coated nanoprisms (See Supplementary Information). Cross-conjugation tests in which pristine (resp. plasma-cleaned) nanoprisms are incubated with NanoGold bearing anti-Au (resp. anti-PVP) antibodies yield nanoprisms with unaltered smooth surface as observed by AFM. This series of conjugation experiments demonstrates the principle of antibody-driven assembly of nanoparticles onto the larger colloids exposing surfaces of the same material used for their selection by phage display. The affinity and selectivity of the selected antibodies for Au vs PVP, PVP vs silica and Au vs silica appears to be high enough to envision the antibody-mediated construction of functional self-assembled plasmonic architectures.

Indeed the close packing of small metallic colloids modifies their starting plasmonic properties such as the emergence of long range coupled modes [31,32]. The Au nanoprisms shown in Figure 4 act as plasmonic resonators sustaining specific higher-order surface plasmon (SP) modes in the visible to infrared range. In particular, the experimental monitoring and the numerical simulations of the surface plasmon local density of states (SP-LDOS) in isolated and coupled nanoprisms reveal that these modes can be spectrally and spatially tuned by shape and size control [33] or by interparticle coupling [30]. Our numeric tool mainly based on the Green dyadic method (GDM) computes accurately the local electromagnetic properties, including SP-LDOS, of 3D metal architectures with complex geometries lying or not on a substrate [19]. We have applied this technique to observe how the presence of additional gold nanoparticles perturb the initial SP-LDOS map of Au nanoprisms as shown in Figure 5 for one particular SP mode of a triangular nanoprism of 700 nm edge length. In the absence of nanoparticles, the spatial distribution of the SP mode centered at 683 nm is strongly concentrated at the center of the prisms with peripheral features along the edges (Figs. 5a-b). Upon random adsorption of a few nanoparticles on the top basal surface (Fig. 5c-d), the SP-LDOS is redistributed with a marked attenuation of the central feature and a concomitant reinforcement of the edge features. Figures 5e and 5f display the SP-LDOS differences and clearly evidence the significant reshaping of the SP modal structure of the initial Au nanoprisms, in spite of the comparatively small diameter of the nanoparticles. The underlying mechanism of this self-assembly-driven SP-LDOS redistribution can be ascribed to spectral detuning of the mode or an alteration of its quality factor in the modified structure.

As a conclusion, the results of our work are threefold. First, we have demonstrated that antibody proteins with affinity for inorganic surfaces can be selected from random libraries without any *a priori* knowledge on the interaction mechanisms between proteins and the targeted solid surface. We confirm that the chemical diversity of current synthetic antibody repertoires is sufficient to obtain strong binders of non-biological targets, the specificity of which matches the one typically encountered in biomolecular antibody/antigen interactions [34]. Even though the intimate interactions between proteins and inorganic solid surfaces are far from being understood in general, our results provide some new insights. Our comprehensive statistical sequence analysis of the selected anti-Au antibodies shows that arginine residues are strongly selected throughout CDR regions where they probably play a major role in the specific gold surface chemisorption. Second, the selected anti-Au and anti-PVP antibodies could be mass-produced and tethered to functionalized Au nanoparticles through their histidine tag without losing their anti-Au or anti-PVP activity. Using both antibodies, we were able to demonstrate the protein-driven self-assembly of gold nanoparticle-nanoprisms superstructures. Finally, we showed that such assembly results in a sizeable modification of the plasmonic properties of the starting Au nanoprisms. Our approach opens the way to colloid-based sensing design opportunities and to the reversible tailoring of the optical response of plasmonic information processing devices based on crystalline metallic nanostructures.

**Materials and Methods**

**Phage display.** The Tomlinson I+J library ([22]) was provided by Source Bioscience and screened against gold powder (purchased from SIGMA) according to the accompanying protocol (http://lifesciences.sourcebioscience.com/media/143421/tomlinsonij.pdf) and to our own protocols [35]. $10^{12}$ to $10^{13}$ phages were prepared by superinfection with helper M13 KO7 phage (GE Healthcare) of exponentially growing TG1 *E. coli* in 2xTY medium, followed by overnight growth at 30°C, PEG (30% w/v)/NaCl (1.5 M) precipitation at 4°C and resuspension in 1 mL PBS. To eliminate polypropylene-binding clones from the libraries, negative selection was performed by incubating the amplified phages suspended in PBS + Tween20 (0.1% w/v, Sigma-Aldrich) + Milk (2% w/v, Marvel fat free) in plain polypropylene 1.5 mL tubes for 1 h at room temperature. The pre-adsorbed library was then incubated on a rocker allowing the phage solution to be shaken evenly for 1 h at room temperature in a polypropylene 1.5 mL tube containing 20 mg gold powder that had been blocked prior to selection with PBS + milk (2% w/v) for 1 hour at room temperature (to avoid non-specific binding). Next, after a 5 s centrifugation on a bench-top centrifuge, the phage solution was removed from the tube, which was subsequently washed 10 times for 5 s with PBS + Tween20 (0.1% w/v) (first round) and 20 times (subsequent rounds) to remove non-specific binders, each time after a 5 s centrifugation. Bound phages were eluted using 1 mL of 100 mM triethylamine (Sigma-Aldrich) for 20 mn and neutralized with 500 μL of Tris buffer (1 M, pH 7.2). Eluted phages were recovered by infection of an excess of exponentially growing TG1 *E. coli* cells (14 mL of a 2xTY culture at O.D.$_{600\ nm}$ = 0.5) for titration and phage preparation for subsequent rounds of selection.

**Expression of selected recombinant antibodies.** 94 selected antibody clones were picked randomly after the 3rd selection round and arrayed into 2 mL 96-well plates. Soluble antibody production was initiated by inoculating 10 μL of an overnight saturated culture (2xTY + Ampicillin + 1% w/v Glucose at 37°C) into 1 mL of 2xTY + Ampicillin + 0.1% w/v Glucose at 37°C. Induction of antibody expression (under the control of the Lac promoter in the Tomlinson library vector) was performed by adding 1 mM IPTG (Isopropyl β-D-1-thiogalactopyranoside, Sigma-Aldrich) to the culture reaching O.D.$_{600\ nm}$ = 0.5, followed by further incubation for 18 hours at 30°C. Soluble antibodies were secreted into the culture medium at a yield of 1 to 10 mg/L. The culture supernatant was then directly used as primary antibody for the ELISA tests. For large-scale production, the expression conditions are scaled up to get a culture of 500 mL. Antibodies are purified on cobalt-ion resin (Co-IDA, Jena Bioscience) through their His6 tag (the Tomlinson library vectors harbor a C-terminal His6 tag sequence downstream of the antibody sequence). Briefly, a 500 mL culture is centrifuged at 10,000xg for 10 minutes and the pellet containing bacteria and debris is discarded. The culture medium is then passed 3 times at 4°C (the procedure is done in a cold room from then on until the end) onto a column containing 1 mL of cobalt-ion resin that had been washed with 10 mL PBS. The column is then washed with 10 mL of PBS + 5 mM imidazole and eluted with 10 mL PBS + 250 mM imidazole (Sigma-Aldrich). 500 μL fractions are collected, dialyzed against PBS, and analyzed by SDS-PAGE. Importantly, the formation of disulfide bonds by sulfhydryl groups in the oxidizing conditions during antibody secretion by bacteria strongly selects against the presence of cysteine residues in the CDR regions of the screened antibodies. Indeed, in spite of the well-known interactions of thiol groups with gold surfaces, we did not identify any cysteine residue at any of the random positions of selected anti-Au antibodies, which may have been eliminated even if such residues had been selected at the phage display stage.

**ELISA binding assay.** A volume of 100 μL of culture medium containing secreted antibodies (or only 2xTY medium for negative controls) was incubated for 1 h at room temperature in 1.5 mL polypropylene tubes (that had been blocked with 1 mL PBS + 0.1% Tween20 + 2% Milk) with or without 1 mg target (Au, Ag, Cu or ZnS powders, Sigma-Aldrich). After 10 s centrifugation on a bench-top centrifuge, the culture medium was discarded and the tube, containing or not the powder target, was rinsed for 15 s in PBS + 0.1% Tween20. Next, a volume of 100 μL of anti-His6 antibody (HIS-1, SIGMA) diluted at 1/1000 in PBS + 0.1% Tween20 + 2% Milk was incubated for 1 h. After rinsing again the tube with PBS + 0.1% Tween20, 100 μL of HRP-conjugated secondary antibodies diluted at 1/3000 in PBS + 0.1% Tween20 + 2% Milk was incubated for 1 h. After rinsing, the ELISA was developed using tetramethylbenzidine (TMB, Sigma-Aldrich). All tests were carried out in duplicate on two separate occasions, and were fully reproducible. Importantly, we noticed that many anti-Au antibodies bind weakly to polystyrene, which precluded the use of polystyrene plates or tubes during binding assay experiments. Previously described polystyrene-binding peptides contain arginine residues, as our anti-Au antibodies do, which may explain our observation.

**Sequencing and antibody sequence analysis.** Phagemid DNA of anti-Au antibody clones, randomly picked antibody clones and anti-PVP antibody clones was purified with miniprep kits from Macherey-Nagel. Sequencing was done at GATC Biotech using the pHEN-SEQ primer, and DNA sequences were translated into amino-acid sequences using TransSeq from EMBOSS. We observed many amber codons (TAG) in the sequences of anti-Au antibodies, all located in the $V_H$-CDR2 regions. These are suppressed and translated as glutamic acid by TG1 *E. coli* (supE genotype) used in our study for phage-display and antibody expression. Amber codons were accordingly translated as glutamic acid. Amino-acid sequences were aligned using the ClustalW 2.1 web-based algorithm (www.ebi.ac.uk/Tools/msa/). Sequence logo was performed using the weblogo (v. 2.8.2) web-based application (http://weblogo.berkeley.edu/logo.cgi). Statistical tests were performed with R software (www.r-project.org/).

**Coupling of anti-Au or anti-PVP antibodies to 5-nm gold nanoparticles.** Anti-Au antibodies produced in this work and anti-polyvinylpyrrolidone (PVP) antibodies produced as described in reference [18]. The as-produced antibody stock solutions are aliquoted and flash-frozen. Just before use, the aliquots are thawed to 4°C. The antibodies are diluted from the initial stock solution (> 1 mg/mL) to 0.01 μM in 20 mM PBS at pH 7 with 150 mM NaCl. The antibody solution are then incubated with the same molar concentration (0.01 μM) of 5 nm Ni-NTA-Nanogold (purchased from NanoProbes Inc.) for 30 minutes at 4°C. The antibody-labeled Nanogold particles are collected by centrifugal filtration with 50K MWCO Amicon filter spun at 5000 rpm for 30 mins.

**Self-assembly of Nanogold onto crystalline gold nanoprisms.** PVP-coated Au nanoprisms were produced as described elsewhere [30]. A 10 μL droplet of nanoprisms suspension is drop-casted onto a clean oxidized (100 nm SiO2) silicon wafer chip and left to dry. For anti-PVP antibody conjugation, the nanoprisms samples are used as such. In order to expose the crystalline gold surface for anti-Au antibody conjugation, substrates with PVP-coated nanoprisms are placed in a $O_2$ plasma cleaner for several 3-minute cycles just before self-assembly [30].

For both anti-PVP and anti-Au antibodies, 10-20 mL of antibody-bearing nanogold solutions (0.01 mM) were added on top of the exposed gold surface of gold nanoprism and incubated for 1 hour at room temperature. The excess liquid was carefully wicked away with filter paper. The substrate was thoroughly rinsed several times with phosphate buffer, deionized water and finally dried using nitrogen gas. The samples were analyzed by AFM (Bruker DI 3000) used in Tapping Mode.

**Supporting Information**.

Supporting Information contains: (1) Anti-Au antibody sequences and (2) TEM images of PVP-coated Au NPR with nanoGold. This material is available free of charge via the Internet at http://pubs.acs.org.

**Corresponding Authors**

* Correspondence should be addressed to C. N. (clement.nizak@espci.fr) or E. D. (dujardin@cemes.fr)

**Author Contributions.** P. J., C. N. and E.D. conceived the experiments. P.J performed the antibody selection and cloning as well as binding specificity assays. P. J., A. S. and C. N. performed the sequence analysis. P. J. and A. S. performed the biochemical modifications of antibodies. J. S. synthesized the nanoprisms. S. N. performed the Nanogold-nanoprism self-assembly and structural characterization (TEM, AFM). C. G. performed GDM calculations. All co-authors contributed to the writing of the article.

**Acknowledgement.** The authors thank P. Minard (IBBMC, University Paris South, France) and B. Dubertret (LPEM laboratory, ESPCI ParisTech, France) for constant support and fruitful discussions. B. Dubertret also provided metal and semi-conductor samples. S. N. and E. D. thank K. L. Gurunatha and A. Thete for technical assistance. E. D. acknowledges the support of the European Research Council (ERC) (contract number ERC–2007-StG Nr 203872 COMOSYEL) and of the Agence Nationale de la Recherche (ANR) (Contract ANR-13-BS10-0007-PlaCoRe). P. J. thanks the Nanoscience Foundation (Grenoble, F) for a PhD fellowship and A. S. acknowledges the support of the SMINGUE Department (Grenoble I University, F) through a postdoctoral fellowship.

**ABBREVIATIONS. AFM**, Atomic Force Microscope. **CDR**, Complementary Determining Region. **Co-IDA**, Cobalt(II)- IminoDiacetic Acid. **ELISA**, Enzyme-Linked ImmunoSorbent Assay. **GDM**, Green dyadic method. **HRP**, horseradish peroxidase. **IPTG**, Isopropyl β-D-1-thiogalactopyranoside. **Ni-NTA**, Nickel(II) NitriloTriacetic Acid. **NMR**, Nuclear Magnetic Resonance. **PBS**, Phosphate-Buffered Saline. **PEG**, poly-ethyleneglycol. **PVP**, poly-vinylpyrrolidone. **scFv**, single chain variable Fragment. **SDS-PAGE**, sodium dodecylsulfate polyacrylamide gel electrophoresis. **SP-LDOS**, Surface Plasmon Local Density of States. **TEM**, Transmission Electron Microscopy.

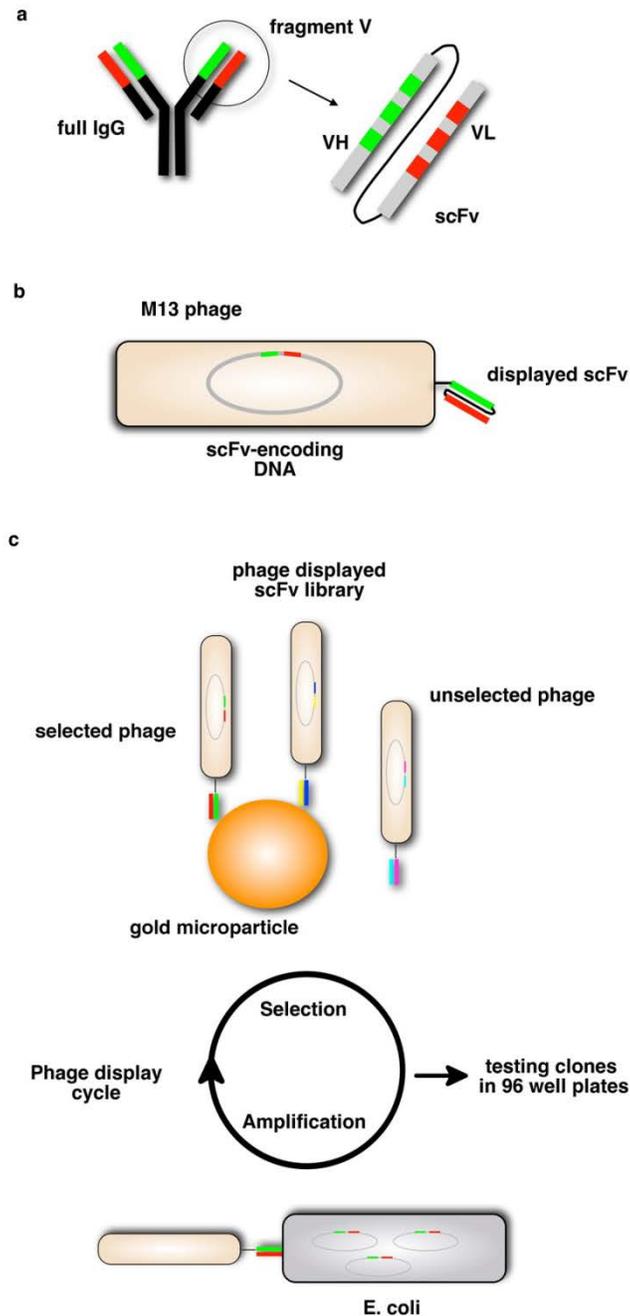

**Figure 1.** Producing anti-Au antibodies by phage-display screening of antibody libraries. (a) The fragment V of antibodies comprises the variable regions of heavy and light chains (respectively $V_H$ and $V_L$), the association of which is generally thought to shape the antigen-binding pocket. $V_H$ and $V_L$ each possess three complementary determining regions (CDR) corresponding to hypervariable sequences (stretches of 5-10 amino-acids, typically 30% of the fragment V sequence overall) that determine binding specificity. Recombinant antibodies of the single-chain fragment V (scFv) format result from the linking of $V_H$ and $V_L$ via a flexible glycine/serine linker. scFvs recapitulate the binding specificity/chemistry of their natural antibody counterparts. (b) Fusion of a scFv-encoding DNA sequence to that of phage surface protein results in the display of this scFv on the phage surface, which provides a physical link between a phage-displayed scFv antibody and its encoding DNA sequence inside the phage capsid. (c) Large, random antibody libraries mimicking natural immune repertoires are screened *in vitro* by phage-display, i.e. alternating cycles of selection against a target antigen and amplification of selected antibody-encoding DNA sequences via phage infection of *E. coli* host bacteria (2 days per cycle). In the Tomlinson I+J antibody library used in the present study, most residues in the CDR regions of scFv antibodies are random while other residues (in the framework regions) are kept constant. During selection, phages that display antibody clones that do not interact strongly with the target are washed away while phages displaying anti-Au antibodies are selected. After generally 3 or 4 cycles of phage-display selection and amplification in bulk, selected antibody clones are arrayed in 96-well plates and tested individually for binding to the target of interest, here gold.

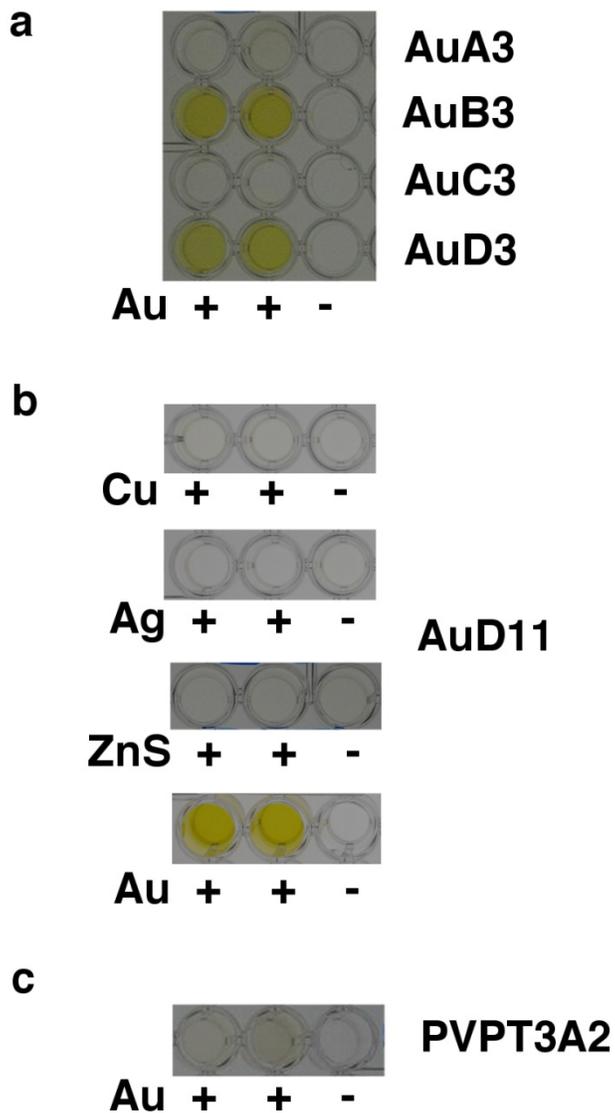

**Figure 2.** Identification of anti-Au antibodies and their binding specificity. (a) 94 antibody clones randomly picked from the 3$^{rd}$ selection round output population were tested for binding to gold using an ELISA binding assay. For each of the 94 antibody clones the culture medium containing secreted antibody was incubated in the presence of gold powder (in duplicate) or not, in a test tube. The intensity of the yellow color relates to the amount of antibody bound to the gold powder. 48% of the 94 clones tested bind to gold, and none of them binds to the test tube. 4 antibody clones are shown here. AuA3 and AuC3 antibody clones do not bind gold; AuB3 and AuD3 antibody clones bind gold. (b) Identified anti-Au antibodies were tested again negative control targets, Ag, Cu, and ZnS. Au was used as positive control target. None of the anti-Au antibodies binds to either Ag, Cu or ZnS. (c) The anti-PVP antibody clone PVPT3A2 isolated from the same library (sharing the same antibody scaffold) was tested for binding to gold. 2xTY medium was used as a negative control, and the AuD11 anti-Au antibody as a positive control (b). The anti-PVP T3A2 antibody did not bind to gold, ruling out non-specific interactions between the antibody scaffold of the Tomlinson I+J library with gold.

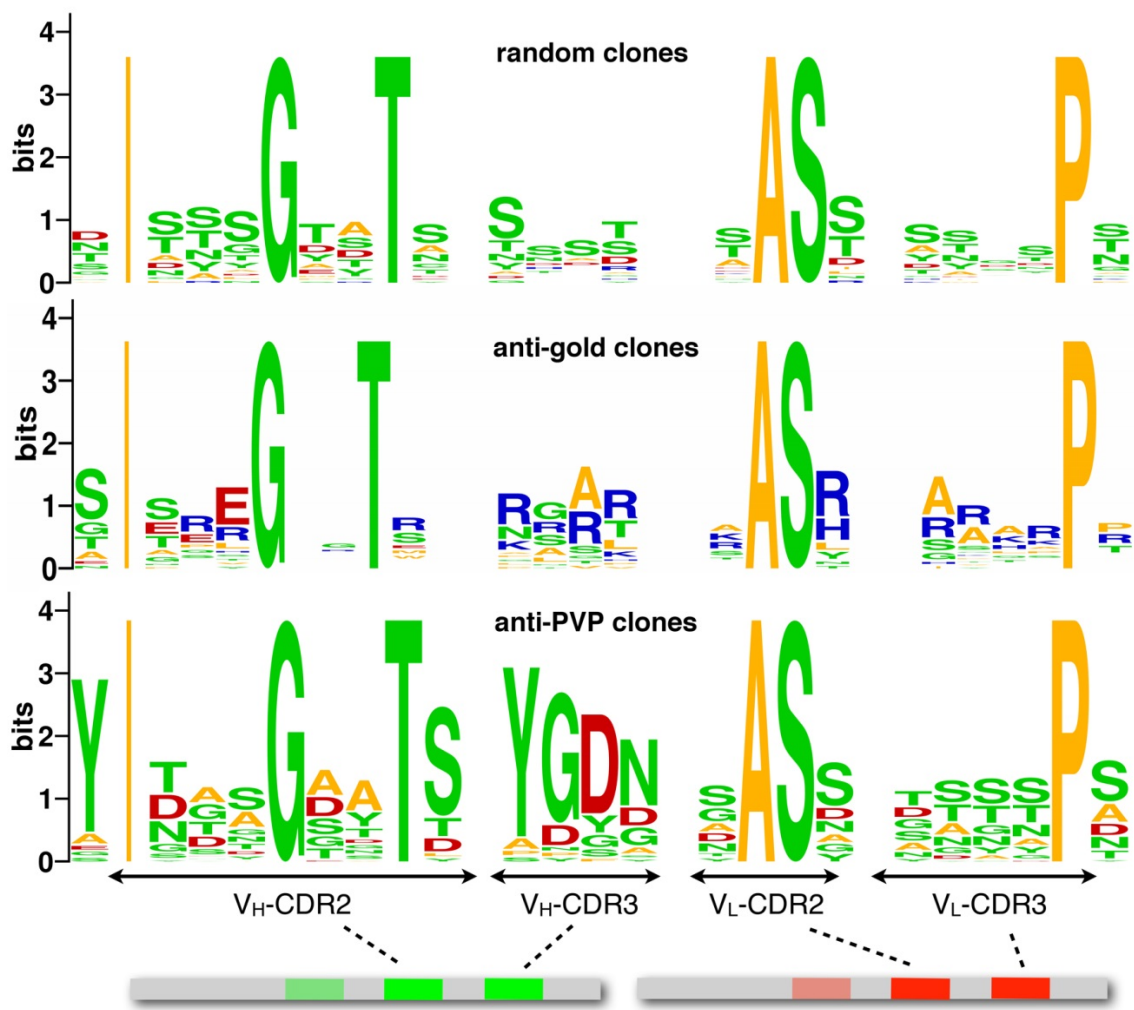

**Figure 3.** Statistical analysis of anti-Au antibody sequences. Sequence logo analysis was performed for 21 distinct anti-Au antibody clones (middle) and 31 distinct antibody clones that were randomly picked from the initial library (top). At each of the 18 random positions, the frequency of each of the 20 amino-acids is computed, and the information content derived. The size of letters representing amino-acids is proportional to their corresponding information content. The 2$^{nd}$, 6$^{th}$ and 8$^{th}$ positions of V$_H$-CDR2, the 2$^{nd}$ and 3$^{rd}$ positions of V$_L$-CDR2 and the 5$^{th}$ position of V$_L$-CDR3 are fixed in the library and the corresponding constant amino-acids (respectively I, G, T, A, S, P) show scores of nearly 4 bits (i.e. the maximum score of any amino-

acid at any random position). Arginine residues (R) are strongly enriched at most random positions across all 4 CDR regions in anti-Au antibodies, in comparison to randomly picked antibody clones from the initial library or anti-PVP antibody clones (bottom) from the same library [18].

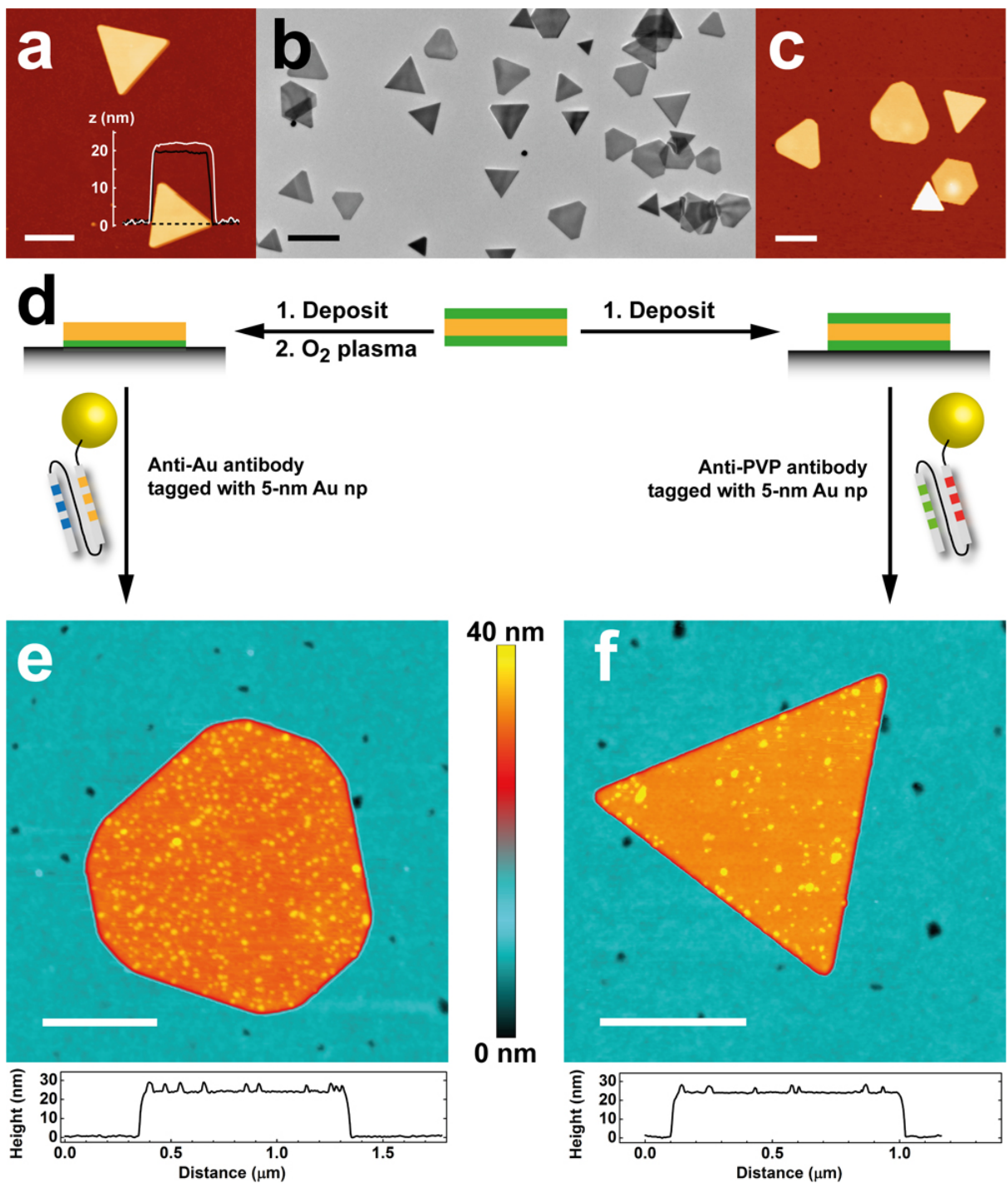

**Figure 4.** Antibody-driven self-assembly of plasmonic Au nanoparticles. (a, c) AFM and (b) TEM images of crystalline Au nanoprisms synthesized as described in reference [30]. The Au nanoprisms that are natively stabilized with PVP are deposited onto oxidized silicon substrate and used (a) with or (c) without a $O_2$ plasma cleaning step that removed the polymer coating. In (a), the inset shows the height profiles before (white) and after (black) O$_2$ plasma cleaning that indicate a reduction of 3-5 nm attributed to the PVP layers. (d) Scheme of the protocol used to expose the bare [111] crystalline facets or the PVP-coated surface (in green) of the Au nanoprisms (in orange) to the selected anti-Au or anti-PVP antibody solutions. (e) AFM image of a Au nanoprism stripped from its PVP coating and thus specifically recognized by the anti-Au antibody tagged with 5-nm Au nanoparticles (yellow dots). Inset: Height profile showing a few 3-5 nm particles on top of the nanoprisms surface. (f) AFM image of a pristine Au nanoprism incubated with anti-PVP antibody tagged with 5-nm Au nanoparticles. The specific attachment to the PVP layer is revealed by the Au nanoparticle adsorption (yellow dots). Inset: Height profile showing a few 3-5 nm particles on top of the nanoprisms surface. (e) and (f) share the same Z color scale bar. Scale bars (a, e, f) 500 nm, (b, c) 1 µm.

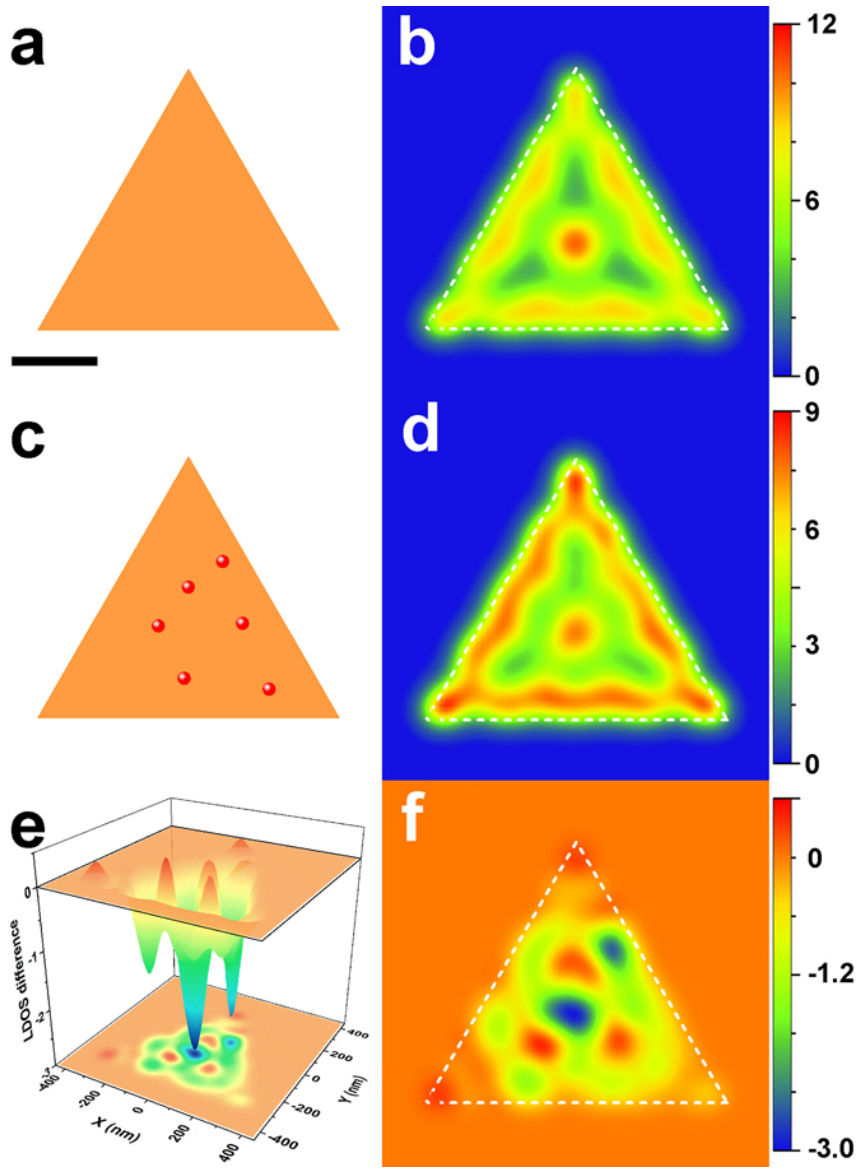

**Figure 5.** Plasmonic modal engineering in self-assembled Au nanoprisms / nanoparticle conjugates. (a) Simulation model of a Au nanoprism with edge length of 700 nm. (b) LDOS map calculated at 683 nm corresponding to a higher order mode (m=4) in the absence of coupled nanoparticles. (c) Simulation model of the Au nanoprism shown in (a) on which six 30 nm diameter spherical Au nanoparticles are randomly adsorbed. (d) LDOS map calculated at 683 nm in the presence of the six nanoparticles. (e, f) 3D and 2D maps of the LDOS difference obtained by subtracting map (b) from map (d). The self-assembly of nanoparticles on the nanoprisms

surface significantly reshape the spatial distribution of the plasmonic LDOS. Common scale bar is 200 nm.

**REFERENCES**

(1)  Dujardin, E.; Mann, S. Synthesis and Assembly of Nanoparticles and Nanostructures Using Bio- Derived Templates. In *Nanobiotechnology II: More Concepts and Applications*; Mirkin, C. A.; Niemeyer, C. M., Eds.; Nanobiotechnology II: More Concepts and Applications: Weinheim, 2007; pp. 39–58.
(2)  Park, S. Y.; Lytton-Jean, A. K. R.; Lee, B.; Weigand, S.; Schatz, G. C.; Mirkin, C. A. DNA-Programmable Nanoparticle Crystallization. *Nature* **2008**, *451*, 553–556.
(3)  Nykypanchuk, D.; Maye, M. M.; van der Lelie, D.; Gang, O. DNA-Guided Crystallization of Colloidal Nanoparticles. *Nature* **2008**, *451*, 549–552.
(4)  Rowan, S. J.; Cantrill, S. J.; Cousins, G. R. L.; Sanders, J. K. M.; Stoddart, J. F. Dynamic Covalent Chemistry. *Angew Chem Int Ed Engl* **2002**, *41*, 898–952.
(5)  Giuseppone, N.; Fuks, G.; Lehn, J.-M. Tunable Fluorene-Based Dynamers Through Constitutional Dynamic Chemistry. *Chemistry* **2006**, *12*, 1723–1735.
(6)  Bradbury, A. R. M.; Sidhu, S.; Dübel, S.; Mccafferty, J. Beyond Natural Antibodies: the Power of in Vitro Display Technologies. *Nature Biotechnology* **2011**, *29*, 245–254.
(7)  Whaley, S. R.; English, D. S.; Hu, E. L.; Barbara, P. F.; Belcher, A. M. Selection of Peptides with Semiconductor Binding Specificity for Directed Nanocrystal Assembly. *Nature* **2000**, *405*, 665–668.
(8)  Mao, C.; Solis, D. J.; Reiss, B. D.; Kottmann, S. T.; Sweeney, R. Y.; Hayhurst, A.; Georgiou, G.; Iverson, B.; Belcher, A. M. Virus-Based Toolkit for the Directed Synthesis of Magnetic and Semiconducting Nanowires. *Science* **2004**, *303*, 213–217.
(9)  Sanghvi, A. B.; Miller, K. P.-H.; Belcher, A. M.; Schmidt, C. E. Biomaterials Functionalization Using a Novel Peptide That Selectively Binds to a Conducting Polymer. *Nature Materials* **2005**, *4*, 496–502.
(10) Tamerler, C.; Oren, E. E.; Duman, M.; Venkatasubramanian, E.; Sarikaya, M. Adsorption Kinetics of an Engineered Gold Binding Peptide by Surface Plasmon Resonance Spectroscopy and a Quartz Crystal Microbalance. *Langmuir* **2006**, *22*, 7712–7718.
(11) Hnilova, M.; Oren, E. E.; Seker, U. O. S.; Wilson, B. R.; Collino, S.; Evans, J. S.; Tamerler, C.; Sarikaya, M. Effect of Molecular Conformations on the Adsorption Behavior of Gold-Binding Peptides. *Langmuir* **2008**, *24*, 12440–12445.
(12) So, C. R.; Tamerler, C.; Sarikaya, M. Adsorption, Diffusion, and Self-Assembly of an Engineered Gold-Binding Peptide on Au(111) Investigated by Atomic Force Microscopy. *Angew Chem Int Ed Engl* **2009**, *48*, 5174–5177.
(13) Rothenstein, D.; Claasen, B.; Omiecienski, B.; Lammel, P.; Bill, J. Isolation of ZnO-Binding 12-Mer Peptides and Determination of Their Binding Epitopes by NMR Spectroscopy. *J Am Chem Soc* **2012**, *134*, 12547–12556.
(14) Lee, Y. J.; Yi, H.; Kim, W.-J.; Kang, K.; Yun, D. S.; Strano, M. S.; Ceder, G.; Belcher, A. M. Fabricating Genetically Engineered High-Power Lithium-Ion Batteries Using Multiple Virus Genes. *Science* **2009**, *324*, 1051–1055.
(15) Dang, X.; Yi, H.; Ham, M.-H.; Qi, J.; Yun, D. S.; Ladewski, R.; Strano, M. S.; Hammond, P. T.; Belcher, A. M. Virus-Templated Self-Assembled Single-Walled Carbon Nanotubes for Highly Efficient Electron Collection in Photovoltaic Devices. *Nature Nanotechnology* **2011**, *6*, 377–384.
(16) Artzy Schnirman, A.; Zahavi, E.; Yeger, H.; Rosenfeld, R.; Benhar, I.; Reiter, Y.; Sivan, U. Antibody Molecules Discriminate Between Crystalline Facets of a Gallium Arsenide Semiconductor. *Nano Lett.* **2006**, *6*, 1870–1874.
(17) Watanabe, H.; Nakanishi, T.; Umetsu, M.; Kumagai, I. Human Anti-Gold Antibodies: Biofunctionalization of Gold Nanoparticles and Surfaces with Anti-Gold Antibodies. *J Biol Chem* **2008**, *283*, 36031–36038.
(18) Soshee, A.; Zürcher, S.; Spencer, N. D.; Halperin, A.; Nizak, C. General in Vitro Method to

Analyze the Interactions of Synthetic Polymers with Human Antibody Repertoires. *Biomacromolecules* **2014**, *15*, 113–121.
(19) Girard, C.; Dujardin, E.; Baffou, G.; Quidant, R. Shaping and Manipulation of Light Fields with Bottom-Up Plasmonic Structures. *New J. Phys.* **2008**, *10*, 105016.
(20) Modi, S.; Nizak, C.; Surana, S.; Halder, S.; Krishnan, Y. Two DNA Nanomachines Map pH Changes Along Intersecting Endocytic Pathways Inside the Same Cell. *Nature Nanotechnology* **2013**, *8*, 459–467.
(21) Modi, S.; Halder, S.; Nizak, C.; Krishnan, Y. Recombinant Antibody Mediated Delivery of Organelle-Specific DNA pH Sensors Along Endocytic Pathways. *Nanoscale* **2013**, *6*, 1144–1152.
(22) de Wildt, R. M.; Mundy, C. R.; Gorick, B. D.; Tomlinson, I. M. Antibody Arrays for High-Throughput Screening of Antibody-Antigen Interactions. *Nature Biotechnology* **2000**, *18*, 989–994.
(23) Mason, J. M.; Müller, K. M.; Arndt, K. M. Protein Engineering Protocols; Methods in molecular biology (Clifton, N.J.), 2007; Vol. 352, pp. 35–70.
(24) Feng, J.; Pandey, R. B.; Berry, R. J.; Farmer, B. L.; Naik, R. R.; Heinz, H. Adsorption Mechanism of Single Amino Acid and Surfactant Molecules to Au {111} Surfaces in Aqueous Solution: Design Rules for Metal-Binding Molecules. *Soft Matter* **2011**, *7*, 2113–2120.
(25) Hoefling, M.; Iori, F.; Corni, S.; Gottschalk, K. E. Interaction of Amino Acids with the Au(111) Surface: Adsorption Free Energies From Molecular Dynamics Simulations. *Langmuir* **2010**, *26*, 8347–8351.
(26) Hoefling, M.; Monti, S.; Corni, S.; Gottschalk, K. E. Interaction of B-Sheet Folds with a Gold Surface. *PLoS ONE* **2011**, *6*, e20925.
(27) Johnston, C. W.; Wyatt, M. A.; Li, X.; Ibrahim, A.; Shuster, J.; Southam, G.; Magarvey, N. A. Gold Biomineralization by a Metallophore From a Gold-Associated Microbe. *Nat Chem Biol* **2013**.
(28) Braun, R.; Sarikaya, M.; Schulten, K. Genetically Engineered Gold-Binding Polypeptides: Structure Prediction and Molecular Dynamics. *Journal of Biomaterials Science, Polymer Edition* **2002**, *13*, 747–757.
(29) Tamerler, C.; Kacar, T.; Sahin, D.; Fong, H.; Sarikaya, M. Genetically Engineered Polypeptides for Inorganics. *Mat Sci Eng C-Bio S* **2007**, *27*, 558–564.
(30) Viarbitskaya, S.; Teulle, A.; Marty, R.; Sharma, J.; Girard, C.; Arbouet, A.; Dujardin, E. Tailoring and Imaging the Plasmonic Local Density of States in Crystalline Nanoprisms. *Nature Materials* **2013**, *12*, 426–432.
(31) Lin, S.; Li, M.; Dujardin, E.; Girard, C.; Mann, S. One-Dimensional Plasmon Coupling by Facile Self-Assembly of Gold Nanoparticles Into Branched Chain Networks. *Adv. Mater.* **2005**, *17*, 2553–2559.
(32) Jägeler-Hoheisel, T.; Cordeiro, J.; Lecarme, O.; Cuche, A.; Girard, C.; Dujardin, E.; Peyrade, D.; Arbouet, A. Plasmonic Shaping in Gold Nanoparticle Three-Dimensional Assemblies. *J. Phys. Chem. C* **2013**, *117*, 23126–23132.
(33) Viarbitskaya, S.; Teulle, A.; Cuche, A.; Sharma, J.; Girard, C.; Dujardin, E.; Arbouet, A. Morphology-Induced Redistribution of Surface Plasmon Modes in Two-Dimensional Crystalline Gold Platelets. *Appl. Phys. Lett.* **2013**, *103*, 131112.
(34) Nizak, C.; Monier, S.; del Nery, E.; Moutel, S.; Goud, B.; Perez, F. Recombinant Antibodies to the Small GTPase Rab6 as Conformation Sensors. *Science* **2003**, *300*, 984–987.
(35) Nizak, C.; Moutel, S.; Goud, B.; Perez, F. Selection and Application of Recombinant Antibodies as Sensors of Rab Protein Conformation. *Meth Enzymol* **2005**, *403*, 135–153.

# Selection of Arginine-Rich Anti-Gold Antibodies Engineered for Plasmonic Colloid Self-Assembly

*Purvi Jain[1], Anandakumar Soshee[1], S Shankara Narayanan[3], Jadab Sharma[3], Christian Girard[3], Erik Dujardin[3*], Clément Nizak[1,2*]*

[1] Laboratory of Interdisciplinary Physics, UMR5588 Grenoble Université 1/CNRS, Grenoble, France.

[2] Laboratory of Biochemistry, UMR 8231 ESPCI ParisTech/CNRS, Paris, France.

[3] CEMES CNRS UPR 8011, 29 rue J. Marvig, 31055 Toulouse Cedex 4, France.

## Supplementary Information

**Table S1:** Sequence alignment table for anti-Au antibodies.

**Figure S2:** TEM of Anti-PVP derivatized Nanogold attached to the surface of Au nanoprisms

**Table S1:** Sequence alignment table for anti-Au antibodies.

```
PC1_D1_E2_D7_C8_H9_C10_C6    SIPEEGSSTT--NSGL--AASY--ASRRPT
PD4                          SIAELGARTR--NLSV--SASY--GSTAPT
PA2                          SIEREGPKTR--NSAL--SASR--AAAKPT
PC2_B12                      SIESYGSRTN--GRVL--SASH--ARHPPT
PG8                          SITTVGEPTR--NGTP--AASR--ARHPPI
PG2                          EINNSGDRTR--RGRR--QASR--RRAKPR
PF5                          GIEARGGKTA--KRRH--AASR--RTAYPR
PB2                          SISERGKPTR--PGRK--DASL--IPLLPP
PA7                          SISETGKITM--AGRK--RASH--RRQRPR
PD5                          SISPRGMETK--LNRR--RASR--HGKRPR
PE6                          SISSRGRETR--FRSR--MASH--RKTGPR
PH4                          AITREGYGTW--KGRR--KASR--ARKQPS
PB7                          AITREGYGTW--KGRR--RASR--RYLKPG
PC4                          SIGVEGLHTS--RSAT--KASN--SRKTPP
PD11                         SIGQHGGVTM--RRAT--HAST--SRKSPP
PD2                          NIEREGTATS--RTAT--RASH--SAHRPP
PB3_A9                       GITPLGRLTE--RLAT--KASL--AANRPP
PG7                          GIAQEGRHTE--RSAT--KASR--GAEAPV
PA4                          TISREGHGTS--RAAR--TASR--AAARPE
PA5_H11                      TISREGHGTS--RAAR--TASR--AAARPA
PC11                         TISREGHGTS--RAAR--AASH--AVQSPQ
```

**Figure S2:** TEM of Anti-PVP derivatized Nanogold attached to the surface of Au nanoprisms

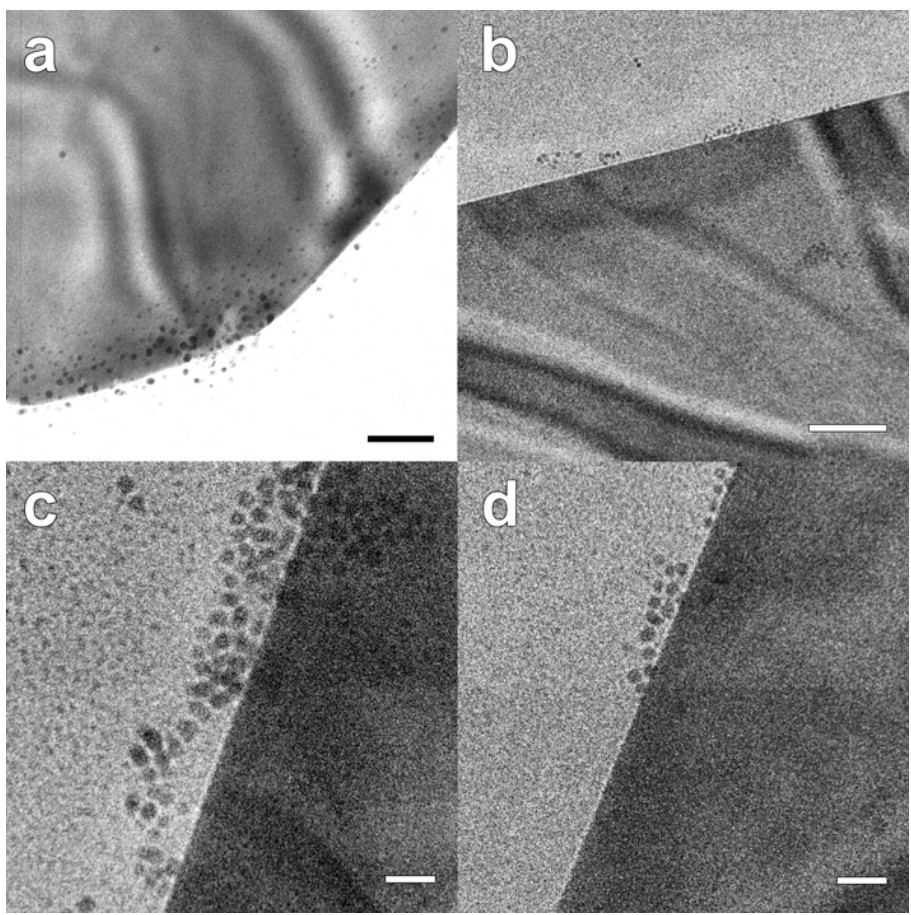

**Figure S2.** Close-up Transmission Electron Micrographs (TEM) of PVP-coated gold nanoprisms decorated with NanoGold nanoparticles labelled with anti-PVP antibody. Scale bars (a) 50 nm, (b) 100nm, (c-d) 20 nm.

The conjugation of Nanogold-labelled anti-PVP antibodies with PVP-coated Au nanoprisms was performed by coupling the anti-PVP antibody to the Ni-NTA Nanogold colloids using the Histidine tag as described in the Experimental Section of the main article. After centrifugal filtration, the Nanogold-labelled antibody was mixed with 20 µL of ultrasonicated nanoprisms suspension and left to incubate. Low speed centrifugation (typically 5000 rpm for 5 minutes) was used to retrieve the nanoprisms that were redispersed in deionized water. The nanoprisms were drop-casted onto Carbon-coated Formwar TEM grids for microscopy analysis. Figure S2 shows a series of micrographs in which the 3-5 nm NanoGold colloids adsorbed on the nanoprisms faces or edges can be observed, therefore confirming that the antibody-PVP interaction did drive the attachment of NanoGold onto the PVP-coated nanoprisms in solution.